\documentclass[11pt,a4paper]{article}

\usepackage[pdftex]{graphics}
\usepackage{jheppub}

\usepackage{amsmath,amssymb,amsfonts}

\usepackage{enumitem}
\usepackage{multirow}

\usepackage{array,booktabs}

\usepackage{slashed}
\usepackage{mathtools}
\usepackage{cancel}
\usepackage{tikz-feynman}
\allowdisplaybreaks
\usepackage{tcolorbox}
\usepackage{float}
\usepackage{tikz,subcaption}
\allowdisplaybreaks
\newcommand{\be}{\begin{eqnarray}}
\newcommand{\ee}{\end{eqnarray}}
\newcommand{\eqa}{\begin{eqnarray}}
\newcommand{\eqae}{\end{eqnarray}}
\newcommand{\nn}{\nonumber}
\newcommand{\bn}{\begin{enumerate}}
\newcommand{\en}{\end{enumerate}}
\newcommand{\bl}{\begin{align}}
\newcommand{\el}{\end{align}}
\newcommand{\eq}{\begin{equation}}
\newcommand{\eqe}{\end{equation}}

\def\identity{{\rlap{1} \hskip 1.6pt \hbox{1}}}
\def\iden{\identity}

\def\CE{{\cal E}}

\def\CH{{\cal H}}

\def\CL{{\cal L}}

\def\CN{{\cal N}}
\def\CO{{\cal O}}

\def\CS{{\cal S}}
\def\CT{{\cal T}}

\def\a{\alpha}
\def\b{\beta}
\def\g{\gamma}

\def\e{\epsilon}
\def\ve{\varepsilon}

\def\th{\theta}

\def\l{\lambda}
\def\m{\mu}
\def\n{\nu}

\def\r{\rho}

\def\s{\sigma}

\def\t{\tau}

\def\w{\omega}

\def\D{\Delta}

\def\S{\Sigma}

\def\half{\frac{1}{2}}

\def\identity{{\rlap{1} \hskip 1.6pt \hbox{1}}}

\newcommand{\bfig}{\begin{figure}}
\newcommand{\efig}{\end{figure}}

\def\la{{\langle}}
\def\ra{{\rangle}}
\def\abs#1{{\left| #1 \right|}}

\def\bl#1\el{\begin{align} #1 \end{align}}
\def\bg#1\eg{\begin{gather} #1 \end{gather}}
\def\bld#1\eld{\begin{aligned} #1 \end{aligned}}
\def\bgd#1\egd{\begin{gathered} #1 \end{gathered}}

\newcommand{\ket}[1]{|{#1}\rangle}

\newcommand{\sbra}[1]{ [{#1} |}

\def\jmath{{j}}
\def\bl#1\el{\begin{align} #1 \end{align}}
\def\bg#1\eg{\begin{gather} #1 \end{gather}}

\def\bld#1\eld{\begin{aligned} #1 \end{aligned}}
\def\bgd#1\egd{\begin{gathered} #1 \end{gathered}}

\newcommand{\eqc}[1]{eq.(\ref{#1})}

\tikzfeynmanset{doublefermion/.style={
/tikz/double,
/tikz/decoration={name=none},
/tikz/postaction={
/tikzfeynman/with arrow=0.5,
}
}
}

\tikzfeynmanset{doubleantifermion/.style={
/tikz/double,
/tikz/decoration={name=none},
/tikz/postaction={
/tikzfeynman/with reversed arrow=0.5,
}
}
}

\title{Spin supplementary condition in quantum field theory: covariant SSC and physical state projection}
\author[1]{Jung-Wook Kim}
\author[1]{Jan Steinhoff}

\affiliation[1]{Max Planck Institute for Gravitational Physics (Albert Einstein Institute),\\ Am M\"uhlenberg 1, Potsdam 14476, Germany}

\emailAdd{jung-wook.kim@aei.mpg.de}
\emailAdd{jan.steinhoff@aei.mpg.de}

\abstract{
The spin supplementary conditions are constraints on spin degrees of freedom in classical relativity which restricts physical degrees of freedom to rotations. It is argued that the equivalent constraints in quantum field theory are the projection conditions on polarisation tensors, which remove timelike/longitudinal polarisations from the physical spectrum. The claim is supported by three examples of massive spinning particles coupled to electromagnetism: Dirac and Proca fields in quantum field theory, and $\CN = 1$ worldline QFT for classical worldline theory. This suggests a resolution to the apparent discrepancy between effective field theory description of massive higher-spin fields~\cite{Bern:2020buy,Bern:2022kto} and post-Newtonian effective field theory of spinning classical particles~\cite{Levi:2015msa}, where the former admits more unfixed parameters compared to the latter; the additional parameters are fixed by projection conditions and therefore are not tunable parameters.
}

\begin{document}
\begin{flushright}
\end{flushright}

\maketitle

\section{Introduction}
The observations of gravitational waves from compact binary coalescence~\cite{LIGOScientific:2016aoc,LIGOScientific:2017vwq} have necessitated a better understanding of compact binary dynamics, attracting practitioners of quantum field theory (QFT) to develop QFT-inspired methods for handling the relativistic two-body problem~\cite{Goldberger:2004jt,Kol:2007bc, Gilmore:2008gq,Foffa:2016rgu,Foffa:2019hrb, Blumlein:2019zku,Foffa:2019rdf,Blumlein:2020pog,Blumlein:2020pyo,Blumlein:2021txe,Almeida:2021xwn,Almeida:2022jrv,Cachazo:2017jef,Luna:2017dtq,Bjerrum-Bohr:2018xdl,Cheung:2018wkq,Caron-Huot:2018ape,Kosower:2018adc,Bern:2019nnu,Bern:2019crd,Brandhuber:2019qpg,AccettulliHuber:2019jqo,KoemansCollado:2019ggb,Cristofoli:2019neg,Cristofoli:2019ewu,Bjerrum-Bohr:2019kec,Cheung:2020gyp,Parra-Martinez:2020dzs,Brandhuber:2021eyq,Bern:2021dqo,DiVecchia:2021bdo,Bjerrum-Bohr:2021vuf,Bjerrum-Bohr:2021din,Bjerrum-Bohr:2021wwt,Bern:2021yeh,Kalin:2020mvi,Kalin:2020fhe,Dlapa:2021npj,Dlapa:2021vgp,Kalin:2022hph,Dlapa:2022lmu,Mogull:2020sak,Jakobsen:2022psy}. An important aspect of relativistic two-body dynamics is the effects of spin, where QFT-inspired methods have proved to be powerful tools for their understanding, as can be inferred from the vast number of recent works on the subject; post-Newtonian effective field theory (PNEFT)~\cite{Levi:2019kgk,Levi:2020kvb,Levi:2020uwu,Levi:2020lfn,Kim:2021rfj,Kim:2022pou,Mandal:2022nty,Kim:2022bwv,Mandal:2022ufb,Levi:2022dqm,Levi:2022rrq}, scattering amplitudes~\cite{Guevara:2017csg,Guevara:2018wpp,Chung:2018kqs,Maybee:2019jus,Guevara:2019fsj,Chung:2019duq,Chung:2020rrz,Bern:2020buy,Kosmopoulos:2021zoq,Aoude:2021oqj,Chen:2021kxt,Aoude:2022trd,Bern:2022kto,Aoude:2022thd,FebresCordero:2022jts}, post-Minkowskian effective field theory~\cite{Liu:2021zxr}, and worldline quantum field theory (WQFT)~\cite{Jakobsen:2021lvp,Jakobsen:2021zvh,Jakobsen:2022fcj,Jakobsen:2022zsx,Wang:2022ntx}. Of course, each approach comes with its own shortcomings; e.g. on-shell-amplitude-based approaches suffer from appearance of unphysical poles~\cite{Arkani-Hamed:2017jhn}, and resolution of their appearance is an active area of research~\cite{Chung:2018kqs,Bautista:2021wfy,Chiodaroli:2021eug,Aoude:2022thd,Aoude:2022trd,Cangemi:2022bew,Bautista:2022wjf}.

Another difficulty is the modelling of multipolar effects due to spin. Spinning compact bodies develop multipole moments due to centrifugal forces, which for the most compact known object of the universe is rather simple: The minimal coupling of ref.\cite{Arkani-Hamed:2017jhn} reproduces the multipolar structures of black holes, both for the gravitational~\cite{Guevara:2018wpp,Chung:2018kqs} and the electromagnetic~\cite{Chung:2019yfs} case. Generic compact spinning objects have richer multipolar structures, however, and their description as point particles in QFT-inspired methods require introduction of Wilson coefficients, which are tunable parameters modelling the strength of the multipoles. Recently, a puzzle regarding the correct number of these parameters were raised in ref.\cite{Bern:2022kto}; the number of tunable parameters for each multipole moment in effective field theory (EFT) of massive higher-spin fields seems to be twice of that in PNEFT, where a spin-induced multipole moment carries only one tunable Wilson coefficient~\cite{Levi:2015msa}. Moreover, it was suggested that the difference in the number of tunable parameters is related to spin supplementary conditions (SSC)~\cite{BernAmp2022}.
Motivated by the confusion over SSCs and their critical role in classical relativity, we initiate a series of studies that investigate the counterparts of SSCs in QFT. This first part of the series is devoted to understanding the covariant SSC in the QFT context using the toy example of a spin-1 massive particle coupled to an electromagnetic field, see the conclusions for future directions.

The SSC 
in classical relativity is a ``gauge'' constraint on antisymmetric rank-2 spin tensor $S^{\m\n}$, usually for massive particles stated in a form of $S^{\m\n} u_\n = 0$ where $u^\m$ is a timelike vector, which determines the centre of the spinning body. The constraint serves to reduce the number of degrees of freedom (DOFs); while the full set of Lorentz generators have $\frac{D(D-1)}{2}$ DOFs, the physical spin only corresponds to spatial rotation generators belonging to the little group which has $\frac{(D-1)(D-2)}{2}$ DOFs.\footnote{In some QFT-based calculations this inspired the definition of the spin tensor as the boost-subtracted Lorentz generator~\cite{Chung:2018kqs,Bautista:2019tdr}. When compared with the contributions from the rotation generators, the contributions from the boost generators can be neglected in the large spin limit ($s \to \infty$) as shown in appendix C of ref.\cite{Chung:2019duq}, therefore it is not necessary to project out the boost generators to define the spin tensor when parity-symmetric spin representations are used. This justifies the definition of the spin tensor definition of ref.\cite{Bern:2020buy}, where the spin tensor and its powers are defined as Lorentz generators and their symmetrised powers sandwiched between polarisation tensors. Note that for fully chiral spin representations boost generators cannot be distinguished from rotation generators~\cite{Chung:2019duq,Aoude:2020onz}, leading to the introduction of GEV in ref.\cite{Guevara:2018wpp}. JWK would like to thank Yilber Fabian Bautista for bringing the discrepancy to attention.} The gap of DOFs is filled by $D-1$ independent constraints imposed by the SSC, which equals the number of boost generators. Consult refs.\cite{1965PhRv..137..188F,Steinhoff:2015ksa} for a review on various SSCs.

In QFT
, spinning particles are classified by the Wigner classification and their spin DOFs are described by polarisation tensors~\cite{Wigner:1939cj,Weinberg:1995mt}. For massive particles in the rest frame, the physical polarisation tensors are purely spacelike tensors transforming irreducibly under the little group $SO(D-1)$,\footnote{We limit our discussion to bosonic case for simplicity.} which is the group of spatial rotations. This condition is also known as the transversality condition, where the contraction of the polarisation tensor with the particle's momentum is constrained to vanish; $p_\m \ve^{\m \n_1 \cdots \n_{s-1}} (p) = 0$.

However, limiting the polarisation tensors to correspond to physical states is not enough, as unphysical states can be excited in quantum processes. The polarisation tensors with timelike components\textemdash which are sometimes referred to as longitudinal polarisations\textemdash do not correspond to physical states, but such tensors can be obtained from the physical polarisation tensors using boost generators. This observation is relevant for vertex rules in Feynman diagrams; the incoming and outgoing states of a single vertex have different momenta, therefore the overlap between physical incoming state and unphysical outgoing state is in general nonzero, leading to excitation of unphysical DOFs that only enters as intermediate states. Therefore, in QFT it is important for the Lagrangian to have special structures such that timelike polarisations become non-dynamical. For example, the higher-spin field Lagrangians constructed in refs.\cite{Singh:1974qz,Singh:1974rc} contain auxiliary fields whose sole purpose is to eliminate unphysical timelike polarisations from the physical spectrum.

Such constraints imposed on higher-spin Lagrangians have an obvious parallel with the SSC; the special structures of the free Lagrangian and the SSC serve to non-dynamicalise DOFs that can be excited through boost generators. The aim of this paper is to argue that the two are different sides of the same coin; the covariant SSC, which non-dynamicalises boost DOFs on the worldline, is equivalent to the projection conditions in QFT, which remove from the physical spectrum the DOFs obtained by action of boost generators on the physical polarisations of spinning fields. The claim is supported by studying three models of spinning particles coupled to electromagnetism; the massive spin-1/2 and spin-1 fields in QFT, and WQFT formalism for classical spinning particles. The advantage of these models is that the impact of SSC on the dynamics can be studied at linear order in spin, which is of lower order than in the corresponding gravitational case. The main ``observable'' is the classical Compton amplitude,\footnote{The absolute square of a scattering amplitude is an observable, but the amplitude itself is not an observable.} where a photon scatters off from a spinning charged particle.

\section{Quantum field theory}
The linear in spin dynamics can be studied in QFT using spin-$\half$ particles~\cite{Holstein:2008sw}. However, Dirac spinors are insufficient for studying the QFT analogue of SSC as their equations of motion (EOM) constrain all DOFs to be physical; the analogue of SSC is expected to be constraints independent of mass-shell constraints that remove unphysical DOFs. On the other hand, the theory of Dirac spinors coupled to electromagnetism can be used as the control group which yields the correct linear-in-spin dynamics, as all DOFs are physical and consideration of SSC is unnecessary.

The ideal model for studying the role of SSC in the QFT context is the Proca action; it is the simplest theory that has enough DOFs to allow unphysical DOFs, which are the timelike polarisations. The Proca action is given as\footnote{Mostly negative sign convention is used for the metric.}
\bl
\bld
\CL &= - \frac{1}{4} G_{\m\n} G^{\m\n} + \frac{m^2}{2} v_\m v^\m = - \frac{1}{2} \partial_\m v_\n \partial^\m v^\n + \frac{1}{2} \partial_\m v_\n \partial^\n v^\m + \frac{m^2}{2} v_\m v^\m \,,
\eld
\el
where the first term, $- \half (\partial_\m v_\n)^2$, combines with the mass term and gives dynamics to the DOFs carried by the vector field $v^\m$. Therefore, it is appropriate to call this term the \emph{kinetic term}. On the other hand, the second term $+ \half \partial_\m v_\n \partial^\n v^\m$ only serves to non-dynamicalise timelike polarisations of $v^\m$, thus it will be called the \emph{projection term}. The role of the projection term can be easily studied through threshold analysis, where fields are only allowed to have time dependence.
\bl
v^\m (t, \vec{x}) \to v^\m (t) && \Rightarrow && \CL = \frac{1}{2} \left[ (\dot{v_i})^2 - \cancel{(\dot{v_0})^2} \right] + \frac{1}{2} \cancel{(\dot{v_0})^2} + \frac{m^2}{2} \left[ v_0^2 - v_i^2 \right] \,.
\el
The projection term cancels the time derivatives of $v_0$ coming from the kinetic term, making the timelike DOF $v_0$ non-dynamical. Note that without the projection term the $v_0$ DOF develops a \emph{negative} kinetic term typical of ghost fields, implying negative norms or violations of unitarity. This is why timelike polarisations are generally considered as unphysical DOFs and special care is taken to remove them from the physical spectrum, e.g. as in models of massive gravity~\cite{deRham:2014zqa}.

The presence(absence) of the projection term corresponds to imposing(lifting) the covariant SSC on the worldline description, which will be demonstrated in the remainder of the paper. The amplitudes are computed in the all-incoming convention and $D = 4$ is implied whenever explicit $D$ dependence is not stated. Details of converting the spinor/polarisation vector expressions to spin vector/tensor expressions can be found in appendix \ref{app:spinop}. The \texttt{xTensor} package of the \texttt{xAct} bundle~\cite{{xAct}} was used in some of the calculations.

\subsection{Action and Feynman rules}
\subsubsection{Spinor QED}
We consider the following Lagrangian for spinor QED
\bl
\bld
\CL &= \bar{\Psi} (i \gamma^\m D_\m - m) \Psi + \frac{i \D g}{8 m} \bar{\Psi} F_{\m\n} \g^\m \g^\n \Psi - \frac{1}{4} (F_{\m\n})^2
\\ &= \bar{\Psi} (i \gamma^\m \partial_\m - m) \Psi + A_\m \bar{\Psi} \g^\m \Psi + \frac{\D g F_{\m\n}}{4 m} \bar{\Psi} \frac{i}{4} [ \g^\m ,  \g^\n ] \Psi - \frac{1}{4} (F_{\m\n})^2 \,,
\eld
\el
where $D_\m = \partial_\m - i A_\m$ is the covariant derivative and $\D g = (g - 2)$ gives the anomalous magnetic moment. The charge of the field $e=-1$ can be determined by identifying $i D_\m$ as the ``mechanical momentum'' $\pi_\m = p_\m - e A_\m$.
\bl
\pi_\m = i D_\m = i \partial_\m + A_\m = p_\m - e A_\m && \Rightarrow && e = -1 \,.
\el
The Feynman rules for this theory are\footnote{The package \texttt{TikZ-Feynman}~\cite{Ellis:2016jkw} was used to draw the diagrams.}
\bn
\item Propagators\footnote{Ordering operators for 2pt functions will be implicit throughout the manuscript.}
\bl
\begin{tikzpicture}[baseline=(a)]
\begin{feynman}
\vertex (a) { $\bar{\Psi}$ } ;
\vertex[right=2cm of a] (b) { $\Psi$ } ;
\diagram*{
(a) -- [fermion, momentum=\(p\)] (b),
};
\end{feynman}
\end{tikzpicture}
&= \langle \Psi(p) \bar{\Psi}(-p) \rangle =  \frac{- i (\slashed{p} + m)}{- p^2 + m^2 - i\e}
\\
\begin{tikzpicture}[baseline=(a)]
\begin{feynman}
\vertex (a) { $A^\m $ } ;
\vertex[right=2cm of a] (b) { $ A^\n $ } ;
\diagram*{
(a) -- [boson, momentum=\(p\)] (b),
};
\end{feynman}
\end{tikzpicture}
&= - i \left[ \frac{- \eta^{\m\n} }{- p^2 - i\e} \right]
\el
\item Vertex
\bl
\begin{tikzpicture}[baseline=(a)]
\begin{feynman}
\vertex (a) { $A^\a $ } ;
\vertex[right=1.5cm of a] (b) ;
\vertex[below right=1.5cm of b] (c) { $\Psi$ } ;
\vertex[above right=1.5cm of b] (d) { $\bar{\Psi}$ } ;
\diagram*{
(a) -- [boson, momentum=\(k_3\)] (b),
(c) -- [fermion, momentum=\(p_1\)] (b),
(d) -- [anti fermion, momentum=\(p_2\)] (b),
};
\end{feynman}
\end{tikzpicture}
&= i \left[
\g^\a + \frac{\D g}{8m}(\slashed{k_3} \g^\a - \g^\a \slashed{k_3})
\right] \label{eq:3vertSpinor}
\el
\en
The only free parameter that can be tuned in this theory is $\D g$.

\subsubsection{Vector QED}
We consider the following Lagrangian for vector QED
\bl
\CL &= - (D_\m v_\n)^\dagger D^\m v^\n + H (D_\m v_\n)^\dagger D^\n v^\m + m^2 v_\m^\dagger v^\m - i (g-H) F^{\m\n} v_\m^\dagger v_\n - \frac{1}{4} (F_{\m\n})^2 \,,\, \label{eq:spin1_action}
\el
where $v_\m$ is a complex vector field, $D_\m = \partial_\m - i A_\m$ is the covariant derivative, and $g$ is the gyromagnetic ratio. The discrete variable $H \in \{ 0,1\}$ parametrises the presence of the projection term; $H = 0$ to switch on dynamical unphysical timelike polarisations, and $H = 1$ to switch them off. The Feynman rules are given as
\bn
\item Propagator
\bl
\begin{tikzpicture}[baseline=(a)]
\begin{feynman}
\vertex (a) { $v^\dagger_\m $ } ;
\vertex[right=2cm of a] (b) { $ v^\n $ } ;
\diagram*{
(a) -- [doublefermion, momentum=\(p\)] (b),
};
\end{feynman}
\end{tikzpicture}
&= - i \left[ \frac{- \delta^\n_\m + \frac{H p^\n p_\m}{m^2 - (1-H) p^2}}{- p^2 + m^2 - i\e} \right] \label{eq:ProcaProp}
\el
\item Vertices
\bl
\begin{tikzpicture}[baseline=(a)]
\begin{feynman}
\vertex (a) { $A^\a $ } ;
\vertex[right=1.5cm of a] (b) ;
\vertex[below right=1.5cm of b] (c) { $v^\m$ } ;
\vertex[above right=1.5cm of b] (d) { $v^\dagger_\n$ } ;
\diagram*{
(a) -- [boson, momentum=\(k_3\)] (b),
(c) -- [doublefermion, momentum=\(p_1\)] (b),
(d) -- [doubleantifermion, momentum=\(p_2\)] (b),
};
\end{feynman}
\end{tikzpicture}
&= i \left[
\begin{aligned}
- (p_1 - p_2)^\a \delta^\m_\n &+ H ( p_{1\n} \eta^{\a\m} - p_2^\m \delta^\a_\n )
\\ &+ (g - H) (-k_{3\n} \eta^{\a\m} + k_3^\m \delta^\a_\n)
\end{aligned}
\right] \label{eq:3vertProca}
\\
\begin{tikzpicture}[baseline=(b)]
\begin{feynman}
\vertex (a1) { $A^\a $ } ;
\vertex[below right=1.5cm of a1] (b) ;
\vertex[below left=1.5cm of b] (a2) { $A_\b$ } ;
\vertex[below right=1.5cm of b] (c) { $v^\m$ } ;
\vertex[above right=1.5cm of b] (d) { $v^\dagger_\n$ } ;
\diagram*{
(a1) -- [boson] (b) -- [boson] (a2),
(c) -- [doublefermion] (b),
(d) -- [doubleantifermion] (b),
};
\end{feynman}
\end{tikzpicture}
&= i \left[ -2 \delta^\a_\b \delta^\m_\n + H (\delta^\a_\n \delta^\m_\b + \eta^{\a\m} \eta_{\b\n}) \right] \label{eq:4vertProca}
\el
\en
with the usual propagator for photons. The propagator of the massive spin-1 field can be obtained by inverting the quadratic action. The $H$ parameter for the propagator and the vertices cannot be independently tuned due to gauge constraints.

Unlike the spinor case, there are two free parameters ($g$ and $H$) for this action. However, only $g$ is continuously tunable and corresponds to $\D g$ of the spinor theory; for generic $H \notin \{ 0,1 \}$ the propagator develops poles away from $p^2 = m^2$, and timelike polarisations become dynamical with masses differing from that of physical polarisations, as can be seen from a threshold analysis.\footnote{The pole of the expression obtained by contracting the propagator \eqc{eq:ProcaProp} with $p^\m p_\n$ corresponds to the mass-shell condition of timelike polarisations. The timelike polarisations with mass-shell conditions differing from that of physical polarisations do not seem to have a corresponding worldline description.}
\bl
v^\m (t, \vec{x}) \to v^\m (t) && \Rightarrow && \CL = \abs{\dot{v_i}}^2 - (1 - H) \abs{\dot{v_0}}^2 + m^2 \left[ \abs{v_0}^2 - \abs{v_i}^2 \right] \,.
\el
The ratio of the coefficients of the time derivative term and the mass term determines the mass-shell conditions, 
and timelike polarisation has the same mass as the physical polarisations only if $H=0$. The timelike polarisation becomes non-dynamical if $H=1$, and the mass-shell condition loses its meaning.

\subsection{Analysis of 3pt amplitudes}
The on-shell 3pt ampliutde for spinor QED is
\bl
\bld
i A_{3,\text{sQED}} &= \begin{tikzpicture}[baseline=(a)]
\begin{feynman}
\vertex (a) ;
\vertex[blob, right=0.9cm of a] (b) {  } ;
\vertex[below right=1.3cm of b] (c) ;
\vertex[above right=1.3cm of b] (d) ;
\diagram*{
(a) -- [boson, momentum=\(k_3\)] (b),
(c) -- [fermion, momentum=\(p\)] (b),
(d) -- [anti fermion, momentum=\(-(p+k_3)\)] (b),
};
\end{feynman}
\end{tikzpicture}
= i \bar{u} (p+k_3) \left[ \slashed{\ve_3} + \frac{\D g}{8m} (\slashed{k_3} \slashed{\ve_3} - \slashed{\ve_3} \slashed{k_3}) \right] u(p) 
\\ &= i \left[ \frac{p \cdot \ve_3}{m} \bar{u}u - \frac{2 + \D g}{4m} i (k_{3\m} \ve_{3\n} - k_{3\n} \ve_{3\m}) \bar{u} \frac{i}{4} [\g^\m , \g^\n ] u \right] \,,
\eld
\el
where the $u$ spinors satisfy the Dirac equation relation $
\slashed{p} u(p) = m u(p)$. In the rest frame of $p^\m$, the first term of the last line can be interpreted as the coupling to the scalar potential $\phi$~\cite{Chung:2018kqs}, while the second term can be interpreted as the coupling to the magnetic field via the spin generator $\S^{\m\n} = \frac{i}{4} [\g^\m , \g^\n]$.\footnote{The identification can be viewed as ``de-quantisation'' of the photon field in the Dyson series expansion of the $S$-matrix. Operationally this corresponds to demoting the annihilation operators and momenta of the photon field to positive frequency components and Fourier modes of the classical background Maxwell field. The idea can be extended to higher order couplings in the background field, e.g. ref.\cite{Kim:2020dif}.} The following identification is used,
\bl
(p \cdot \ve_3) \leftrightarrow m A_0 = m \phi \,,\,~~ k_3^i \ve_3^j - k_3^j \ve_3^i \leftrightarrow i (\partial^i A^j - \partial^j A^i) = - i \ve^{ijk} B^k \,, \label{eq:3ptHamRel}
\el
where $\phi$ is the scalar potential and $\vec{B}$ is the magnetic field. Converting the spin generator to the spin tensor $\bar{u} \S^{\m\n} u = \bar{u} \mathbb{S}^{\m\n} u$, the 3pt amplitude can be written as
\bl
\bld
A_{3,\text{sQED}} &= \bar{u} (p+k_3) \left[ \phi \iden - \frac{2 + \D g}{4m} \e^{ijk} B^k \mathbb{S}^{ij} \right] u(p)
\\ &= - \bar{u} (p+k_3) \left[ e \phi \iden - \frac{e(2 + \D g)}{2m} B^k \mathbb{S}^{k} \right] u(p) \,,
\eld \label{eq:3ptSpinorHamRel}
\el
where the charge $e = -1$ of the particle has been restored in the last line. Comparing the expression with the Zeeman Hamiltonian,
\bl
H_Z = Q \phi - \vec{\m} \cdot \vec{B} \,,\,~~ \vec{\m} = \frac{Qg}{2m} \vec{S} \,,
\el
the combination $g = 2 + \D g$ is identified as the gyromagnetic ratio.\footnote{The Dyson series expression for the $S$-matrix is $S = \iden + i T = \CT \exp( - i \int H dt)$, and amplitudes are defined as the matrix elements of the $T$-matrix. This absorbs the extra sign between $A_3$ and $H_Z$.}

The on-shell 3pt amplitude for vector QED is
\bl
\bld
i A_{3,\text{vQED}} &=
\begin{tikzpicture}[baseline=(a)]
\begin{feynman}
\vertex (a) ;
\vertex[blob, right=0.9cm of a] (b) {  } ;
\vertex[below right=1.3cm of b] (c) ;
\vertex[above right=1.3cm of b] (d) ;
\diagram*{
(a) -- [boson, momentum=\(k_3\)] (b),
(c) -- [doublefermion, momentum=\(p_1\)] (b),
(d) -- [doubleantifermion, momentum=\(p_2\)] (b),
};
\end{feynman}
\end{tikzpicture}
\\ &= - i \left[ 2 (p_1 \cdot \ve_3) (\ve_2 \cdot \ve_1) + g \left\{ (\ve_2 \cdot k_3)(\ve_3 \cdot \ve_1) - (\ve_2 \cdot \ve_3) (k_3 \cdot \ve_1) \right\} \right] \,. \label{eq:3ptEMspin1}
\eld
\el
Although the vertex rule \eqc{eq:3vertProca} depends on $H$, the dependence drops out in the on-shell amplitude. Note that the Ward identity for the massive spin-1 particle $\ve_1 \to p_1$ is satisfied only when $g$ is set to $g=2$.\footnote{This requirement is similar to but different from higher-spin gauge symmetry proposed in ref.\cite{Cangemi:2022bew}; the higher spin gauge symmetry is intended to work for off-shell currents.} Using the identification \eqc{eq:3ptHamRel}, the 3pt amplitude can be written as
\bl
\bld
A_{3,\text{vQED}} &= - 2m \ve_2^i \left[ \phi (-\delta^{ij}) + \frac{g}{2m} B^k (-i \ve^{ijk}) \right] \ve_1^j
\\ &= - 2m \ve_2^i \left[ e \phi \delta^{ij} - \frac{eg}{2m} B^k (\mathbb{S}^k)^{ij} \right] \ve_1^j \,,
\eld \label{eq:3ptVecHamRel}
\el
where unit charge $e = -1$ has been restored to the final expression and the matrix element expression for the spin operator $(\mathbb{S}^k)^{ij} = -i \ve^{kij}$ has been used. This form of the 3pt amplitude  has an obvious parallel to the spinor expression \eqc{eq:3ptSpinorHamRel}, and the parameter $g$ in the vector QED is identified as the gyromagnetic ratio. 

\subsection{Compton amplitudes in the classical limit} \label{sec:QFTCompton}

For the helicity-preserving configuration, the polarisation vectors of the external photon states are set as
\bl
\ve_1^\m = \ve_{1+}^\m = \frac{\sbra{1} \bar{\s}^\m \ket{4}}{\sqrt{2} \la 14 \ra} \,,\, \ve_4^\m = \ve_{4-}^\m = \frac{\sbra{1} \bar{\s}^\m \ket{4}}{\sqrt{2} [ 41 ]} \,,
\el
where spinor definitions follow the conventions of ref.\cite{Chung:2018kqs} and $k_4$ is the outgoing particle, i.e. $k_4^0 < 0$. For the helicity-flipping configuration, the polarisation vectors of the external photon states are set as
\bl
\ve_1^\m = \ve_{1+}^\m = \frac{\sbra{4} \bar{\s}^\m \ket{1}}{\sqrt{2} [ 14 ]} \,,\, \ve_4^\m = \ve_{4+}^\m = \frac{\sbra{1} \bar{\s}^\m \ket{4}}{\sqrt{2} [ 41 ]} \,.
\el
The Mandelstam variables are defined as $s = (p_2 + k_1)^2 = m^2 + 2 m \w \hbar$ and $t = (k_1 + k_4)^2 = q^2 \hbar^2$, where $\w$ is the frequency of the (incoming) photon in the rest frame\footnote{The incoming momentum ($p_2^\m$) defines the rest frame.} of the massive spinning particle. Conventions for the spin vector can be found in appendix \ref{app:spinop}. The classical limit is taken using the $\hbar$ counting~\cite{Kosower:2018adc,Chung:2019duq}
\bl
k_1^\m \to \hbar k_1^\m \,,\, k_4^\m \to \hbar k_4^\m \,,\, \mathbb{S}^\m \to \hbar^{-1} \mathbb{S}^\m \,,
\el
together with the $\hbar$ counting of Mandelstam variables and expanding the expression to leading order in $\hbar$. The scattering angle $\th$ is defined by the relation $q^2 = - 4 \w^2 \sin^2 (\th / 2)$.

\subsubsection{Classical Compton amplitude for spinor QED}
The Compton amplitude is obtained by adding two diagrams.
\bl
i A_{4,\text{sQED}} &=
\begin{tikzpicture}[baseline=(b)]
\begin{feynman}
\vertex (a1) ;
\vertex[blob, below right=1.2cm of a1] (b) { };
\vertex[below left=1.5cm of b] (a2) ;
\vertex[below right=1.5cm of b] (c) ;
\vertex[above right=1.5cm of b] (d) ;
\diagram*{
(a2) -- [boson, momentum=\(k_1\)] (b),
(a1) -- [boson, momentum=\(k_4\)] (b),
(c) -- [fermion, momentum=\(p_2\)] (b),
(d) -- [anti fermion, momentum=\(p_3\)] (b),
};
\end{feynman}
\end{tikzpicture}
=
\begin{tikzpicture}[baseline=(b3)]
\begin{feynman}
\vertex (a1) ;
\vertex[below right=1cm of a1] (b1) ;
\vertex[below=1cm of b1] (b2) ;
\vertex[below=0.5cm of b1] (b3) ;
\vertex[below left=1cm of b2] (a2) ;
\vertex[below right=1cm of b2] (c) ;
\vertex[above right=1cm of b1] (d) ;
\diagram*{
(a2) -- [boson] (b2),
(a1) -- [boson] (b1),
(c) -- [fermion] (b2) -- [fermion] (b1) -- [fermion] (d),
};
\end{feynman}
\end{tikzpicture}
+
\begin{tikzpicture}[baseline=(b3)]
\begin{feynman}
\vertex (a1) ;
\vertex[below right=1cm of a1] (b1) ;
\vertex[below=1cm of b1] (b2) ;
\vertex[below=0.5cm of b1] (b3) ;
\vertex[right=0.7cm of b1] (b1p) ;
\vertex[right=0.7cm of b2] (b2p) ;
\vertex[below left=1cm of b2] (a2) ;
\vertex[below right=1cm of b2p] (c) ;
\vertex[above right=1cm of b1p] (d) ;
\diagram*{
(a2) -- [boson] (b1p),
(a1) -- [boson] (b2p),
(c) -- [fermion] (b2p) -- [fermion] (b1p) -- [fermion] (d),
};
\end{feynman}
\end{tikzpicture}
\el
For the helicity-preserving configuration, the Compton amplitude is
\begin{subequations}
\bl
\bld
i A_{4,\text{sQED}}^{+-} &= \bar{u}(-p_3) \left[ - \frac{i (4 \w^2 + q^2)}{4 m \w^2} + \frac{i[(g-2)^2 \w^2 - g q^2]}{8 m^2 \w^2} [(k_1 - k_4) \cdot \mathbb{S}] \right.
\\ &\phantom{=asdfasdfasdfasdfasdfasdfasdf} \left. + \frac{(4 + g^2)}{8 m^3 \w} \e_{\m\n\l\s}k_1^\m p_2^\n k_4^\l \mathbb{S}^\s \right] u(p_2)
\\  &= \bar{u}(-p_3) \left[ - \frac{i \cos^2(\th/2)}{m} + \frac{i[g^2 - 2 g (1 + \cos \th) + 4]}{8 m^2} [(k_1 - k_4) \cdot \mathbb{S}] \right.
\\ &\phantom{=asdfasdfasdfasdfasdfasdfasdf} \left. + \frac{(4 + g^2)}{8 m^3 \w} \e_{\m\n\l\s}k_1^\m p_2^\n k_4^\l \mathbb{S}^\s \right] u(p_2) \,,
\eld \label{eq:sQEDCompHelPres}
\el
where $g = 2 + \D g$. For the helicity-flipping configuration, the Compton amplitude is
\bl
\bld
i A_{4,\text{sQED}}^{++} &= \bar{u}(-p_3) \left[ - \frac{i q^2}{4 m \w^2} - \frac{i ((g^2 -4) \w^2 - g q^2)}{8 m^2 \w^2} [(k_1 + k_4) \cdot \mathbb{S}] \right.
\\ &\phantom{=asdfasdfasdfasdfasdfasdfasdf} \left. - \frac{(g^2 - 4) }{8 m^3 \w} \e_{\m\n\l\s}k_1^\m p_2^\n k_4^\l \mathbb{S}^\s \right] u(p_2)
\\  &= \bar{u}(-p_3) \left[ \frac{i \sin^2(\th/2)}{m} + \frac{i[g^2 + 2 g (1 - \cos \th) - 4]}{8 m^2} [(k_1 + k_4) \cdot \mathbb{S}] \right.
\\ &\phantom{=asdfasdfasdfasdfasdfasdfasdf} \left. + \frac{(g^2 - 4)}{8 m^3 \w} \e_{\m\n\l\s}k_1^\m p_2^\n k_4^\l \mathbb{S}^\s \right] u(p_2) \,.
\eld \label{eq:sQEDCompHelFlip}
\el
\label{eq:sQEDComp}
\end{subequations}

\subsubsection{Classical Compton amplitude for vector QED}
The Compton amplitude is obtained by adding three diagrams.
\bl
i A_{4,\text{vQED}} &=
\begin{tikzpicture}[baseline=(b)]
\begin{feynman}
\vertex (a1) ;
\vertex[blob, below right=1.2cm of a1] (b) { };
\vertex[below left=1.5cm of b] (a2) ;
\vertex[below right=1.5cm of b] (c) ;
\vertex[above right=1.5cm of b] (d) ;
\diagram*{
(a2) -- [boson, momentum=\(k_1\)] (b),
(a1) -- [boson, momentum=\(k_4\)] (b),
(c) -- [doublefermion, momentum=\(p_2\)] (b),
(d) -- [doubleantifermion, momentum=\(p_3\)] (b),
};
\end{feynman}
\end{tikzpicture}
=
\begin{tikzpicture}[baseline=(b3)]
\begin{feynman}
\vertex (a1) ;
\vertex[below right=1cm of a1] (b1) ;
\vertex[below=1cm of b1] (b2) ;
\vertex[below=0.5cm of b1] (b3) ;
\vertex[below left=1cm of b2] (a2) ;
\vertex[below right=1cm of b2] (c) ;
\vertex[above right=1cm of b1] (d) ;
\diagram*{
(a2) -- [boson] (b2),
(a1) -- [boson] (b1),
(c) -- [doublefermion] (b2) -- [doublefermion] (b1) -- [doublefermion] (d),
};
\end{feynman}
\end{tikzpicture}
+
\begin{tikzpicture}[baseline=(b3)]
\begin{feynman}
\vertex (a1) ;
\vertex[below right=1cm of a1] (b1) ;
\vertex[below=1cm of b1] (b2) ;
\vertex[below=0.5cm of b1] (b3) ;
\vertex[right=0.7cm of b1] (b1p) ;
\vertex[right=0.7cm of b2] (b2p) ;
\vertex[below left=1cm of b2] (a2) ;
\vertex[below right=1cm of b2p] (c) ;
\vertex[above right=1cm of b1p] (d) ;
\diagram*{
(a2) -- [boson] (b1p),
(a1) -- [boson] (b2p),
(c) -- [doublefermion] (b2p) -- [doublefermion] (b1p) -- [doublefermion] (d),
};
\end{feynman}
\end{tikzpicture}
+
\begin{tikzpicture}[baseline=(b)]
\begin{feynman}
\vertex (a1) ;
\vertex[below right=1.5cm of a1] (b);
\vertex[below left=1.5cm of b] (a2) ;
\vertex[below right=1.5cm of b] (c) ;
\vertex[above right=1.5cm of b] (d) ;
\diagram*{
(a2) -- [boson] (b),
(a1) -- [boson] (b),
(c) -- [doublefermion] (b) -- [doublefermion] (d),
};
\end{feynman}
\end{tikzpicture}
\el
While the Compton amplitude can be analysed up to $\CO(S^2)$ order in spin, the Compton amplitude will be truncated to linear order as the results will only be compared up to this order in spin. For the helicity-preserving configuration without propagating timelike polarisations ($H=1$), the Compton amplitude is
\begin{subequations}
\bl
\bld
i A_{4,\text{vQED}}^{H=1\,,+-} &= - \frac{i (4 \w^2 + q^2)}{2 \w^2} + \frac{i[(g-2)^2 \w^2 - g q^2]}{4 m \w^2} [(k_1 - k_4) \cdot \mathbb{S}]
\\ &\phantom{=asdfasdfasdfasdfasdf} + \frac{(4 + g^2)}{4 m^2 \w} \e_{\m\n\l\s}k_1^\m p_2^\n k_4^\l \mathbb{S}^\s
\\ &= - 2 i \cos^2 (\th/2) + \frac{i[g^2 - 2 g (1 + \cos \th ) + 4 ]}{4 m} [(k_1 - k_4) \cdot \mathbb{S}]
\\ &\phantom{=asdfasdfasdfasdfasdf} + \frac{(4 + g^2)}{4 m^2 \w} \e_{\m\n\l\s}k_1^\m p_2^\n k_4^\l \mathbb{S}^\s \,,
\eld \label{eq:PQEDCompHelPres}
\el
where $g$ is the gyromagnetic ratio. This expression matches the spinor case \eqc{eq:sQEDCompHelPres} up to normalisation of the spinors $\bar{u} u = 2m$. For the helicity-flipping configuration, it is
\bl
\bld
i A_{4,\text{vQED}}^{H=1\,,++} &= - \frac{i q^2}{2 \w^2} - \frac{i [(g^2 -4) \w^2 - g q^2]}{4 m \w^2} [(k_1 + k_4) \cdot \mathbb{S}] - \frac{(g^2 - 4) }{4 m^2 \w} \e_{\m\n\l\s}k_1^\m p_2^\n k_4^\l \mathbb{S}^\s
\\ &= 2 i \sin^2 (\th/2) - \frac{i [g^2 + 2g(1-\cos \th) -4]}{4 m} [(k_1 + k_4) \cdot \mathbb{S}]
\\ &\phantom{=asdfasdfasdfasdfasdf} - \frac{(g^2 - 4) }{4 m^2 \w} \e_{\m\n\l\s}k_1^\m p_2^\n k_4^\l \mathbb{S}^\s \,,
\eld \label{eq:PQEDCompHelFlip}
\el
which also matches the spinor case \eqc{eq:sQEDCompHelFlip}.
\label{eq:PQEDComp}
\end{subequations}

Interestingly, the effect of projection condition \emph{does not vanish} in the classical limit,
\begin{subequations}
\bl
i A_{4,\text{vQED}}^{H=1\,,+-} - i A_{4,\text{vQED}}^{H=0\,,+-} &= \frac{i (g-2)^2 \left[ m \w \{ (k_1 - k_4) \cdot \mathbb{S} \} - i \e_{\m\n\l\s}k_1^\m p_2^\n k_4^\l \mathbb{S}^\s \right]}{4 m^2 \w} \,, \label{eq:helpresCompClassDiff}
\\ i A_{4,\text{vQED}}^{H=1\,,++} - i A_{4,\text{vQED}}^{H=0\,,++} &= \frac{i (g-2)^2 \left[ m \w \{(k_1 + k_4) \cdot \mathbb{S}\} - i \e_{\m\n\l\s}k_1^\m p_2^\n k_4^\l \mathbb{S}^\s \right]}{4 m^2 \w} \,, \label{eq:helflipCompClassDiff}
\el
and the difference is directly proportional to $(\D g)^2 = (g-2)^2$. \label{eq:CompClassDiff}
\end{subequations}
This is another indication that $g = 2$ is a special coupling~\cite{Ferrara:1992yc}, which is also the coupling of Kerr-Newman black holes to electromagnetic fields~\cite{Carter:1968rr} and the gyromagnetic ratio for the minimal coupling amplitudes in the sense of ref.\cite{Arkani-Hamed:2017jhn}.
It may be of interest to the reader that the differences \eqc{eq:CompClassDiff} contribute to the classical limit of one-loop integral coefficients.

\section{Worldline description from worldline QFT}
In the worldline description of spinning particles, the particles are treated as classical particles and the SSC can be explicitly imposed on the worldline action. A convenient worldline description for studying effects of SSC is the worldline quantum field theory (WQFT) formalism~\cite{Mogull:2020sak,Jakobsen:2021lvp}, where the preservation of covariant SSC on the worldline can be traded for the preservation of supersymmetry (SUSY) on the worldline~\cite{Jakobsen:2021zvh}. This simplifies the constraint equations as can be seen by writing the worldline translation generator as $\CH$ and writing the preservation conditions as
\bl
\{ S^{\m\n} u_\n , \CH \} \simeq 0 \quad \left( D-1\text{ conditions, SSC} \right) && \text{v.s.} && \{ Q , \CH \} \simeq 0 \quad \left( 1\text{ condition, SUSY} \right) \nn
\el
where $\simeq$ is an equality up to the spin order of interest. The reduction in the number of constraint equations shows the advantage of the WQFT description.

To compare the QFT amplitude results with worldline-based calculations, a WQFT action with (broken) $\CN = 1$ SUSY is constructed. Since it is known that $(g-2)$ can be viewed as a measure of SUSY breaking~\cite{Ferrara:1974wb}, and considering the fact that $C_E$ in the gravitational case could be tuned by a ``soft'' SUSY breaking operator which allowed approximate conservation of SUSY charges, we expect two sets of operators that contribute to $(g-2)$ in the electromagnetic case; the ``soft'' breaking term which approximately preserves the SUSY charges, and the ``hard'' breaking term which do not preserve the SUSY charges in any sense. The expectation is that the ``soft''(``hard'') breaking corresponds to $H=1$($H=0$) of vector QED, since SUSY preservation is interpreted as SSC preservation~\cite{Jakobsen:2021zvh}.

\subsection{Construction of WQFT action}
Following the treatement of ref.\cite{Jakobsen:2021zvh}, the supercharge is written as\footnote{See also ref.\cite{Comberiati:2022cpm} for a QCD generalisation.}
\bl
Q = \psi \cdot \pi + \th m \,,
\el
where the Poisson brackets are defined by\footnote{Following ref.\cite{Jakobsen:2021zvh}, the Grassmann variable for the compactified direction is denoted as $\th$, and the momentum $p_{\varphi}$ dual to the compactified coordinate $\varphi$ is fixed as a constant $p_{\varphi} = m$. The momentum variable $p_\r$ for the Grassmann variable $\r$ is defined using the right derivative $p_\r := - \left. \frac{\delta S}{\delta \dot{\r}} \right|_R$. The Poisson brackets for Grassmann variables should be understood as Dirac brackets instead; see e.g. appendix \ref{app:Grass_Quant}.}
\bl
\{ x^\m, p_\n \} = \delta^\m_\n \,,\, \{ \psi^\m, \psi^\n \} = - i \eta^{\m\n} \,,\, \{ \th, \th \} = + i \,, \label{eq:PB}
\el
and the ``mechanical momentum'' $\pi_\m$ is defined as
\bl
\pi_\m = p_\m - q A_\m \Rightarrow \{ \pi_\m , \pi_\n \} 
= q F_{\m\n} \,.
\el
leading to the Hamiltonian
\bl
-2i\CH = \{ Q, Q \} = - i \left[ \pi \cdot \pi - m^2 + i q F_{\m\n} \psi^\m \psi^\n \right] \,, \label{eq:WQFTnopertHam}
\el
and the action 
\bl
\bld
S &= - \int d\t \left[ p_\m \dot{x}^\m + m \dot{\varphi} + \frac{i}{2} \psi \cdot \dot{\psi} - \frac{i}{2} \th \dot{\th} - e \CH - i \chi Q \right]
\\ &= - \int d\t \left[ p_\m \dot{x}^\m + m \dot{\varphi} + \frac{i}{2} \psi \cdot \dot{\psi} - \frac{i}{2} \th \dot{\th} - \frac{e}{2} \left( \pi^2 - m^2 + i q F_{\m\n} \psi^\m \psi^\n \right) - i \chi \left( \psi \cdot \pi + \th m \right) \right]
\eld \label{eq:wlact_orig}
\el
where $e$ is the einbein and $\chi$ is the gravitino. The variation in $e$ yields the mass-shell constraint $\CH = 0$ and the variation in $\chi$ yields the SUSY conservation constraint. The mass-shell constraint can be solved to compute the energy $\CE$ of the particle
\bl
\bld
\CH &= 0 = (\CE - q\phi)^2 - (\vec{p} - q \vec{A})^2 - m^2 - q F_{\m\n} S^{\m\n} \,.
\\ \therefore \CE &= q \phi + \sqrt{m^2 + q F_{\m\n} S^{\m\n} + (\vec{p} - q \vec{A})^2} \simeq q \phi + m - \frac{q \vec{S}}{m} \cdot \vec{B} + \frac{(\vec{p} - q \vec{A})^2}{2m} \,,
\eld \label{eq:WQFTnopertEn}
\el
where $S^{\m\n} = - i \psi^\m \psi^\n$ was adopted and the positive root has been taken. This is the energy of a particle with gyromagnetic ratio $g=2$ on a background electromagnetic field.

This spin tensor definition is justified since the algebra satisfied by $S^{\m\n} = - i \psi^\m \psi^\n$ is identical to the one by orbital angular momentum tensor $L^{\m\n} = x^\m p^\n - x^\n p^\m$.
\bl
\bld
\{ L^{\m\n}, L^{\a\b} \} &= \eta^{\m\a} L^{\n\b} - \eta^{\n\a} L^{\m\b} + \eta^{\m\b} L^{\a\n} - \eta^{\n\b} L^{\a\m} \,,
\\ \{ S^{\m\n}, S^{\a\b} \} &= \eta^{\m\a} S^{\n\b} - \eta^{\n\a} S^{\m\b} + \eta^{\m\b} S^{\a\n} - \eta^{\n\b} S^{\a\m} \,.
\eld
\el
While the algebra of $S^{\m\n}$ is canonical which implies the canonical SSC~\cite{Steinhoff:2011sya}, the dynamics of $S^{\m\n}$ is \emph{different} from that expected for the canonical SSC and follows the covariant SSC instead~\cite{Jakobsen:2021zvh}. 
Since $m$ is constant and $\varphi$ decouples from the rest of the variables, the compactified direction kinetic term $m \dot{\varphi}$ can be dropped when studying the dynamics.

The expectation for $g \neq 2$ (which corresponds to non-vanishing ``finite-size'' effects $C_E \neq 0$ in the gravitational context) is that SUSY is broken~\cite{Ferrara:1974wb} and there are two sets of worldline operators responsible for SUSY breaking; the one that preserves the SSC (the ``soft'' SUSY breaking), and the one that does not preserve the SSC (the ``hard'' SUSY breaking).\footnote{It is known that a theory with $g \neq 2$ \emph{cannot} be double copied to a gravitational theory because the coupling of spin to gravity is universal~\cite{Holstein:2006ry,Goldberger:2017ogt,Chung:2018kqs}. Therefore this term is irrelevant for double-copy construction of integrands for gravitational dynamics~\cite{Comberiati:2022cpm}.} The soft breaking term is expected to break SUSY at $\CO(\psi^3)$ or higher orders in Grassmann variables, while the hard breaking term is expected to break SUSY at $\CO(\psi^2)$ or lower orders. Starting from the ansatz for $(g-2)$ interaction $\CH_{g-2}$,
\bl
\CH_{g-2} &= c_1 F_{\m\n} \psi^\m \psi^\n + c_2 \frac{(\pi \cdot \psi) F_{\m\n} \pi^\m \psi^\n}{\pi^2} \,,
\el
the SUSY variation on $\CH_{g-2}$ yields;
\bl
\bld
\{Q , \CH_{g-2} \} &= c_1 \{ \psi \cdot \pi , F_{\m\n} \psi^\m \psi^\n \} + c_2 \{ \psi \cdot \pi , \frac{(\psi \cdot \pi) F_{\m\n} \pi^\m \psi^\n}{\pi^2} \}
\\ &= {- 2 i c_1} F_{\m\n} \pi^\m \psi^\n + c_2 \frac{\{ \psi \cdot \pi , (\psi \cdot \pi) F_{\m\n} \pi^\m \psi^\n \}}{\pi^2} + \CO(\psi^3)
\\ &= - i ({2 c_1 + c_2}) F_{\m\n} \pi^\m \psi^\n + \CO(\psi^3) \,,
\eld
\el
which singles out the soft SUSY breaking combination as $2c_1 + c_2 = 0$. This combination can be written using the projector $P^\m_{\n}$ as
\bl
F_{\m\n} \psi^\m \psi^\n - 2 \frac{(\pi \cdot \psi) F_{\m\n} \pi^\m \psi^\n}{\pi^2} = F_{\a\b} P^\a_\m P^\b_\n \psi^\m \psi^\n \,,\, P^\m_\n = \delta^\m_\n - \frac{\pi^\m \pi_\n}{\pi^2} \,,
\el
where the projector projects onto the transverse space orthogonal to $\pi^\m$. To fix the normalisation for $\CH_{g-2}$, consider $\pi^\m$ to be directed along the time direction in the rest frame of the particle. Using the relation $S^{\m\n} = - i \psi^\m \psi^\n$ this combination can be rewritten as an anomalous Zeeman coupling term
\bl
- i \frac{(g-2) q}{4 m} F_{\a\b} P^\a_\m P^\b_\n \psi^\m \psi^\n \simeq \frac{(g-2) q}{4m} F_{\a\b} P^\a_\m P^\b_\n S^{\m\n} \simeq \frac{(g-2) q}{2m} (-\vec{B}) \cdot \vec{S} \,,
\el
fixing the natural normalisation as $c_1 = i (g-2) q / 4$, or\footnote{A factor of $(-m)$ is multiplied to the worldline Hamiltonian; see \eqc{eq:WQFTnopertHam} and \eqc{eq:WQFTnopertEn}.}
\bl
\CH_{g-2} &= i \frac{(g-2) q}{4} F_{\a\b} P^\a_\m P^\b_\n \psi^\m \psi^\n = i \frac{(g-2) q}{4} \left[ F_{\m\n} \psi^\m \psi^\n - \frac{2 (\pi \cdot \psi) F_{\m\n} \pi^\m \psi^\n}{\pi^2} \right] \,.
\el
The free parameter that can be tuned without changing the gyromagnetic ratio is $c_2$,\footnote{The $c_2$ tunes electric-dipole-like coupling $\propto \vec{p} \cdot \vec{E}$.} which is parametrised by $h$ in parallel to the $H$ parameter of the QFT case. The perturbation $\D \CH$ to the Hamiltonian $\CH$ appearing in \eqc{eq:wlact_orig} is parametrised as
\bl
\D \CH = \CH_{\D g,h} = i \frac{\textcolor{red}{\D g} q}{4} \left[ F_{\m\n} \psi^\m \psi^\n - \textcolor{red}{h} \frac{2 (\pi \cdot \psi) F_{\m\n} \pi^\m \psi^\n}{\pi^2} \right] \,. \label{eq:delta_Ham}
\el
where tunable parameters $\D g = (g-2)$ and $h$ were highlighted in red for discernability. The expectation is that $h = 0$ corresponds to $H = 0$ and $h = 1$ to $H = 1$; SSC preservation on the worldline ($h=1$) is expected to match non-dynamical timelike polarisations for the field theory ($H=1$). A similar argument can be made for $h=0$ and $H=0$, which can be found in appendix \ref{app:h0H0}.\footnote{A key difference between the parameters $h$ and $H$ is that while $H$ is forced to be discrete by mass-shell conditions, there does not seem to be an analogous condition forcing $h$ to be discrete. The authors would like to thank the anonymous referee for pointing this out.} The substitution of $\CH \to \CH + \CH_{\D g,h}$ in \eqc{eq:wlact_orig} yields
\bl
S &= - \int d\t \left[ p_\m \dot{x}^\m + m \dot{\varphi} + \frac{i}{2} \psi \cdot \dot{\psi} - \frac{i}{2} \th \dot{\th} - e (\CH + \CH_{\D g,h}) - i \chi Q \right] \,. \label{eq:wlact_mod}
\el
The form of the action suitable for calculations is obtained by inserting the equations of motion from the variation by $p^\m$,
\bl
\bld
- \frac{\delta S}{\delta p_\m} &= 0 = \dot{x}^\m - e \left( \pi^\m - \frac{i \D g h q}{2 \pi^2} \left[ \left\{ \psi^\m  - \frac{2 \pi^\m (\pi \cdot \psi)}{\pi^2} \right\} F_{\a\b} \pi^\a \psi^\b + (\pi \cdot \psi) F^{\m\a} \psi_\a \right] \right) - i \chi \psi^\m \,,
\eld \nn
\el
which are solved perturbatively to $\CO(\psi^3)$, i.e.
\bl
\bld
\pi^\m &= e^{-1} (\dot{x}^\m - i \chi \psi^\m) + \frac{i \D g h q}{2 \pi^2} \left[ \left\{ \psi^\m  - \frac{2 \pi^\m (\pi \cdot \psi)}{\pi^2} \right\} F_{\a\b} \pi^\a \psi^\b + (\pi \cdot \psi) F^{\m\a} \psi_\a \right]
\\ &= e^{-1} (\dot{x}^\m - i \chi \psi^\m) + \frac{i \D g h e q}{2 \dot{x}^2} \left[ \left\{ \psi^\m  - \frac{2 \dot{x}^\m (\dot{x} \cdot \psi)}{\dot{x}^2} \right\} F_{\a\b} \dot{x}^\a \psi^\b + (\dot{x} \cdot \psi) F^{\m\a} \psi_\a \right] + \CO(\psi^3) \,.
\eld \nn
\el
When studying linear-in-spin dynamics, terms of order $\CO(\psi^3)$ and higher can be neglected. The action \eqc{eq:wlact_mod} now becomes
\bl
\bld
S &= - \int d\t \left[ \frac{\dot{x}^2}{2e} + q A \cdot \dot{x} + \frac{i}{2} \psi \cdot \dot{\psi} - \frac{i}{2} \th \dot{\th}  - i \chi \left( \frac{\dot{x} \cdot \psi}{e} + m \th \right) \right.
\\ & \phantom{=asdf} \left. \phantom{} + \frac{e}{2} \left( m^2 - \frac{i (2 + \D g) q}{2} F_{\m\n} \psi^\m \psi^\n + i \D g h q \frac{(\dot{x} \cdot \psi) F_{\m\n} \dot{x}^\m \psi^\n}{\dot{x}^2} \right) \right] + \CO(\psi^3) \,.
\eld
\el
Under gauge-fixing by $\chi = 0$ and $e = m^{-1}$, together with the rescaling $\psi^\m \to \sqrt{m} \psi^\m$, the action becomes
\bl
\bld
S &= - \int d\t \left[ \frac{m}{2} \dot{x}^2 + q A \cdot \dot{x} + \frac{i}{2} m \psi \cdot \dot{\psi} - \frac{i (2 + \D g) q}{4} F_{\m\n} \psi^\m \psi^\n \right.
\\ &\phantom{=asdfasdfasdf} \left. \phantom{asdfasdfasdf} + \frac{i \D g h q (\dot{x} \cdot \psi) F_{\m\n} \dot{x}^\m \psi^\n}{2 \dot{x}^2} 
+ \cdots
\right] 
\,,
\eld \label{eq:WQFT_act_fin}
\el
where elipsis denotes terms irrelevant for Feynman rules. The rescaling of $\psi^\m$ rescales the spin tensor definition as $S^{\m\n} = - i m \psi^\m \psi^\n$. 

\subsection{Derivation of Feynman rules}
To derive the Feynman rules, the worldline fields are decomposed into their background values and their fluctuations.
\bl
\bld
x^\m &= b^\m + v^\m \t + z^\m (\t) \,,
\\ \psi^\m &= \Psi^\m + \psi'^\m (\t) \,.
\eld \label{eq:bg_exp}
\el
Whenever an expression quadratic in the background Grassmann field $\Psi^\m$ appears, it will be written using the background spin tensor $\CS^{\m\n} = - i m \Psi^\m \Psi^\n$ instead.

The boundary conditions on the fluctuation fields $z^\m$ and $\psi'^\m$ determine the boundary conditions of their propagators. The boundary condition for Feynman propagators corresponds to the symmetric boundary condition for the fluctuation fields~\cite{Jakobsen:2021zvh}. The worldline fluctuation fields and the gauge field are expanded in terms of positive frequency modes\footnote{In the classical theory, this sign choice guarantees analyticity of (causal) worldline fluctuations on the upper half plane of $\w$ space, provided that worldline fluctuations are polynomially bounded in $\t$.}
\bl
\bld
z^\m (\t) &= \int_\w e^{- i \w \t} z^\m (\w) & && \leftrightarrow & & && z^\m (\w) &= \int_\t e^{+ i \w \t} z^\m (\t) \,,
\\ \psi'^\m (\t) &= \int_\w e^{- i \w \t} \psi'^\m (\w) & && \leftrightarrow & & && \psi'^\m (\w) &= \int_\t e^{+ i \w \t} \psi'^\m (\t) \,,
\\ A^\m (x) &= \int_k e^{- i k \cdot x} A^\m (k) & && \leftrightarrow & & && A^\m (k) &= \int_x e^{+ i k \cdot x} A^\m (x) \,,
\eld \label{eq:mode_exp}
\el
which relates them to annihilation operators and \emph{incoming} convention for the momenta,\footnote{Note that \emph{outgoing} convention was used for the same expansion in ref.\cite{Jakobsen:2021zvh}.} where the measure for each space is defined as
\bl
\int_\w := \int \frac{d\w}{2\pi} \,,\qquad \int_k := \int \frac{d^D k}{(2\pi)^D} \,,\qquad \int_\t := \int d\t \,,\qquad \int_x := \int d^D x \,.
\el
The Feynman rules can be read out by expanding the worldline action \eqc{eq:WQFT_act_fin} in fluctuation fields \eqc{eq:bg_exp} and inserting the mode expansions \eqc{eq:mode_exp}. 

\subsubsection{Worldline propagators}
The propagators for worldline fluctuation fields are obtained from the quadratic terms of the free action, which is given as
\bl
S_{\text{free}} &= - \frac{m}{2} \int_\w \left\{ [v^2 - 2i\w v \cdot z(\w) + \w^2 z(-\w) \cdot z(\w)] + \w [\Psi \cdot \psi'(\w) + \psi'(-\w) \cdot \psi'(\w)] \right\} \,. \nn
\el
The terms linear in fluctuations are proportional to the classical equations of motion satisfied by the background fields, and therefore can be dropped. The quadratic terms determine the propagators;\footnote{The possible sign factor from ordering of the two-point function has been absorbed by the functional integral measure.}
\bl
\begin{tikzpicture}[baseline=(a)]
\begin{feynman}
\vertex (a) { $z^\m (-\w) $ } ;
\vertex[right=3cm of a] (b) { $ z^\n (\w) $ } ;
\diagram*{
(a) -- [momentum=\(\w\)] (b),
};
\end{feynman}
\end{tikzpicture}
&= \la z^\n (\w) z^\m (-\w) \ra = - i \frac{\eta^{\m\n}}{m \w^2}
\\
\begin{tikzpicture}[baseline=(a)]
\begin{feynman}
\vertex (a) { $\psi'^\m(-\w)$ } ;
\vertex[right=3cm of a] (b) { $ \psi'^\n(\w) $ } ;
\diagram*{
(a) -- [scalar, momentum=\(\w\)] (b),
};
\end{feynman}
\end{tikzpicture}
&= \la \psi'^\n (\w) \psi'^\m (-\w) \ra = - i \frac{\eta^{\m\n}}{
m \w}
\el
where the $i\e$ prescription has been dropped for simplicity. The $i\e$ prescription can be restored depending on the choice of boundary conditions for the propagators, e.g. $\w \to \w + i\e$ for causal propagation along the arrow. The propagators are related to the time domain 2pt functions by the following Fourier transform.
\bl
\la f(\t) f(0) \ra = \int_\w e^{- i \w \t} \la f(\w) f(-\w) \ra \,.
\el

\subsubsection{Interaction vertices}
The terms linear in $A^\m$ generate the interaction vertices. The terms without worldline fluctuation fields give the background coupling
\bl
\bld
i S_{\text{int}}^{\text{bg}} &= \int_k e^{- i k \cdot b} \bar{\delta} \left( k \cdot v \right) \left[ - i q (A \cdot v) - \frac{(2 + \D g)q}{2} \frac{k_\m A_\n \CS^{\m\n}}{m} \right.
\\ &\phantom{=asdfasdf} \left. \phantom{asdfasdfasdf} + \frac{\D g h q}{2 v^2} \frac{(k_\m A_\n - k_\n A_\m) v_\a v^\m \CS^{\a\n}}{m} \right] \,.
\eld
\el
When covariant SSC is imposed on the background spin tensor ($\CS^{\m\n}v_\n = 0$), the last term vanishes and the background coupling becomes independent of the parameter $h$.

The linear coupling to $z^\m$ fluctuations is given as
\bl
\bld
i S_{\text{int}}^{z} &= \int_k \int_\w e^{- i k \cdot b} \bar{\delta} \left( k \cdot v + \w \right) \left[ - q \left\{ (A \cdot v) (k \cdot z) + \w (A \cdot z) \right\} \phantom{\frac{(2 + \D g) q}{2}\frac{k_\m A_\n \CS^{\m\n}}{m}
} \right.
\\ &\phantom{=asdfasdf} + i \frac{(2 + \D g) q}{2}\frac{k_\m A_\n \CS^{\m\n}}{m} (k \cdot z) - i \frac{\D g h q}{2 v^2} \frac{\w (k_\m A_\n - k_\n A_\m) v^\m z_\a \CS^{\a\n}}{m}
\\ &\phantom{=asdfasdf} \left. - i \frac{\D g h q}{2 v^2} \left[ (k \cdot z) v^\m - 2 \w \frac{(v \cdot z)}{v^2} v^\m + \w z^\m \right] \frac{(k_\m A_\n - k_\n A_\m) v_\a \CS^{\a\n}}{m} \right] \,,
\eld \label{eq:zFeynRule}
\el
where the last line vanishes when covariant SSC is imposed. It is clear from the expression that propagating $z^\m$ fluctuations cannot contribute to $\CO( (\D g)^2 \CS^1)$ interactions at $\CO(q^2)$ order, where $q$ is the charge. Note that the vertex rule still depends on the parameter $h$ even if covariant SSC is imposed.

The linear coupling to $\psi'^\m$ fluctuations is given as
\bl
\bld
i S_{\text{int}}^{\psi'} &= \int_k \int_\w e^{- i k \cdot b} \bar{\delta} \left( k \cdot v + \w \right) (k_\m A_\n - k_\n A_\m) \left[ i \frac{(2 + \D g) q}{2} \Psi^\m \psi'^\n - i \frac{\D g h q}{2 v^2} \Psi^\m v^\n (v \cdot \psi') \right.
\\ &\phantom{=asdfasdfasdfasdfasdfasdfasdfasdfasdfasdf} \left. - i \frac{\D g h q}{2 v^2} (v \cdot \Psi) v^\m \psi'^\n \right]
\eld \label{eq:psiFeynRule}
\el
where the $(v \cdot \Psi)$ dependent term has been separated out to the last line. The covariant SSC for the background spin tensor can be alternatively expressed as $(v \cdot \Psi) = 0$~\cite{Jakobsen:2021zvh}. Imposing the SSC removes the last line from the vertex rule, although the $h$ dependence persists similar to the $z^\m$ case.

\subsection{The Compton amplitude and comparison with vector QED}
The scattering of a photon off the spinning particle consists of two terms; worldline propagation of $\psi'$ fluctuations and worldline propagation of $z$ fluctuations. The conditions $v^2 = 1$ and $(v \cdot \Psi) = 0$ are imposed to simplify the calculations, leading to the expression
\bl
\bld
& e^{- i (k_1 + k_4) \cdot b} \bar{\delta} \left( (k_1 + k_4) \cdot v \right)
\\ & \phantom{\int} \times - i \frac{2 q^2}{m} \left[ \frac{(\ve_1 \cdot v) (\ve_4 \cdot v) (k_1 \cdot k_4)}{(k_1 \cdot v)^2} + \frac{(\ve_1 \cdot v) (\ve_4 \cdot k_1) - (\ve_4 \cdot v)(\ve_1 \cdot k_4)}{k_1 \cdot v} - (\ve_1 \cdot \ve_4) \right.
\\ & \phantom{\int asdf} - i \left( \frac{2 + \D g}{2} \right) \frac{(k_1 \cdot \CS \cdot \ve_1)}{m} \left( \frac{(\ve_4 \cdot v)(k_1 \cdot k_4)}{(k_1 \cdot v)^2} + \frac{k_1 \cdot \ve_4}{k_1 \cdot v} \right)
\\ & \phantom{\int asdf} - i \left( \frac{2 + \D g}{2} \right) \frac{(k_4 \cdot \CS \cdot \ve_4)}{m} \left( \frac{(\ve_1 \cdot v)(k_1 \cdot k_4)}{(k_1 \cdot v)^2} - \frac{k_4 \cdot \ve_1}{k_1 \cdot v} \right)
\\ & \phantom{\int asdf} - i \frac{(k_1 \cdot \CS \cdot k_4)}{m} \left( \frac{(2+\D g)^2 (\ve_1 \cdot \ve_4)}{4 (k_1 \cdot v)} - \frac{(\D g)^2 (2h - h^2)}{4} \frac{(\ve_1 \cdot v)(\ve_4 \cdot v)}{k_1 \cdot v} \right)
\\ & \phantom{\int asdf} - i \frac{(\ve_1 \cdot \CS \cdot \ve_4)}{m} \left( \frac{(2+\D g)^2 (k_1 \cdot k_4)}{4 (k_1 \cdot v)} + \frac{(\D g)^2 (2h - h^2)}{4} (k_1 \cdot v) \right)
\\ & \phantom{\int asdf} + i \frac{(k_1 \cdot \CS \cdot \ve_4)}{m} \left( \frac{(2+\D g)^2 (\ve_1 \cdot k_4)}{4 (k_1 \cdot v)} + \frac{(\D g)^2 (2h - h^2)}{4} (\ve_1 \cdot v) \right)
\\ & \phantom{\int asdf} \left. + i \frac{(\ve_1 \cdot \CS \cdot k_4)}{m} \left( \frac{(2+\D g)^2 (k_1 \cdot \ve_4)}{4 (k_1 \cdot v)} - \frac{(\D g)^2 (2h - h^2)}{4} (\ve_4 \cdot v) \right) \right] \,.
\eld
\el
The notation $(p \cdot \CS \cdot q) = p_\m \CS^{\m\n} q_\n$ was used above. To compare with the amplitude results \eqc{eq:PQEDComp}, the WQFT Compton amplitude is defined as the above expression without the impact parameter space transformation factors $e^{- i (k_1 + k_4) \cdot b} \bar{\delta} \left( (k_1 + k_4) \cdot v \right)$. Normalising to unit charge ($q^2 = 1$) and using the polarisation vectors defined in sec.~\ref{sec:QFTCompton}, the Compton amplitudes with $h=1$ become
\begin{subequations}
\bl
i A_{\text{WQFT}}^{h=1\,,+-} &= - \frac{2 i \cos^2 (\th/2)}{m} + \frac{i[g^2 - 2 g (1 + \cos \th ) + 4 ]}{4 m^2} [(k_1 - k_4) \cdot \CS] \nn
\\ &\phantom{=asdfasdfasdfasdf} + \frac{(4 + g^2)}{4 m^2 \w} \e_{\m\n\l\s}k_1^\m k_4^\n \CS^\l v^\s \,,
\\ i A_{\text{WQFT}}^{h=1\,,++} &= \frac{2 i \sin^2 (\th/2)}{m} - \frac{i (g^2 + 2g(1-\cos \th) -4)}{4 m^2} [(k_1 + k_4) \cdot \CS] \nn
\\ &\phantom{=asdfasdfasdfasdf} - \frac{(g^2 - 4) }{4 m^2 \w} \e_{\m\n\l\s}k_1^\m k_4^\n \CS^\l v^\s \,,
\el
where $\hbar$ counting is done with $(k_1 \cdot v) = \w \hbar$ and $(k_1 \cdot k_4) = - 2 \w^2 \hbar^2 \sin^2 (\th/2)$, the spin tensor is converted to the spin vector $\CS^\m$ by the relation $\CS^{\m\n} = - \e^{\m\n\a\b} v_\a \CS_{\b}$, and the superscript $+-$($++$) denotes helicity preserving(flipping) configuration.
\label{eq:WQFTCompton}
\end{subequations}
These expression match with \eqc{eq:PQEDComp} up to an overall $m^{-1}$ factor\footnote{This factor can be absorbed by rescaling the delta constraint $m^{-1} \bar{\delta}(k \cdot v) = \bar{\delta} (k \cdot mv) = \bar{\delta} (k \cdot p)$.} under the identification $p_2^\m = m v^\m$. Moreover, the differences between $h=1$ and $h=0$ also match with \eqc{eq:CompClassDiff}.
\begin{subequations}
\bl
i A_{\text{WQFT}}^{h=1\,,+-} - i A_{\text{WQFT}}^{h=0\,,+-} &= \frac{i (g-2)^2 \left[ \w \{(k_1 - k_4) \cdot \CS \} - i \e_{\m\n\l\s}k_1^\m k_4^\n \CS^\l v^\s \right]}{4 m^2 \w} \,,
\\ i A_{\text{WQFT}}^{h=1\,,++} - i A_{\text{WQFT}}^{h=0\,,++} &= \frac{i (g-2)^2 \left[ \w \{(k_1 + k_4) \cdot \CS\} - i \e_{\m\n\l\s}k_1^\m k_4^\n \CS^\l v^\s \right]}{4 m^2 \w} \,,
\el
which is a strong indication that covariant SSC of the worldline theory, parametrised by the $h$ coefficient, is related to the projection conditions in QFT, parametrised by the $H$ coefficient.
\label{eq:WQFTComptonDiff}
\end{subequations}

\section{Conclusion}
The match in the classical limit between WQFT Compton amplitudes \eqc{eq:WQFTCompton} and vector QED Compton amplitudes \eqc{eq:PQEDComp}, especially the match between the effect of SSC breaking term in WQFT amplitudes \eqc{eq:WQFTComptonDiff} and the effect of removing the projection term in vector QED amplitudes \eqc{eq:CompClassDiff}, supports the view that imposing covariant SSC on the worldline description of spinning particles is equivalent to removing unphysical timelike polarisations from the physical spectrum of higher-spin fields. Interestingly, on-shell methods do not suffer from this problem since unphysical polarisations are removed from the get-go, although the results generally suffer from unphysical intermediate states in the form of unphysical poles which need to be removed by other means~\cite{Chung:2018kqs,Chiodaroli:2021eug,Aoude:2022thd,Aoude:2022trd}. Furthermore, let us stress that the unphysical DOFs have an impact on classical observables if they are not projected out at the level of the Lagrangian.\footnote{Or if the propagators do not have proper projectors, when the QFT is understood in the Weinbergian sense~\cite{Weinberg:1996kw} where the Lagrangian is simply a means of obtaining consistent interaction vertex rules. Maintaining gauge invariance can become nontrivial in this approach; the $H$ parameter dependence in vector QED is an example of projectors affecting the vertex Feynman rules.}

The view that covariant SSC and physical state projection conditions are equivalent has immediate consequences. One consequence is that it suggests a possible resolution to the puzzle raised in EFT description of massive higher-spin fields for binary dynamics~\cite{Bern:2022kto}. In this construction, two families of operators modelling spin-multipole effects were found to contribute independently to classical dynamics. Whereas one family of operators\textemdash the $C$-type operators\textemdash have corresponding counterparts in the worldline theory of spinning particles introduced in ref.\cite{Levi:2015msa}, the other family of operators\textemdash the $H$-type operators\textemdash do not have corresponding counterparts, which seems to imply an apparent doubling of the number of tunable parameters. The view suggests that the mismatch is due to not imposing the transversality conditions on the massive higher-spin fields~\cite{Bern:2020buy,Bern:2022kto}, and that removing the unphysical timelike polarisations from the physical spectrum will induce additional constraints on unfixed parameters such that the number of tunable parameters will match that of PNEFT. Another consequence is that the view offers a guiding principle for constructing WQFT action of spinning particles with higher order spin-induced multipole moments; the operators are chosen so that SUSY is broken as ``softly'' as possible.

One method of removing unphysical timelike polarisations from the physical spectrum is to introduce St\"uckelberg fields~\cite{deRham:2014zqa}. Therefore it is likely that the higher-spin gauge symmetry of ref.\cite{Cangemi:2022bew}, which uses St\"uckelberg fields for gauge symmetry, automatically implements the physical state projection conditions. It would be interesting to check if the doubling of tunable parameters observed in ref.\cite{Bern:2022kto} is resolved when higher-spin gauge symmetry is imposed, i.e. the higher-spin gauge symmetry allows at most one free parameter to be tuned at each spin-multipole moment order. Moreover, explicitly studying the higher-spin case would also clarify if the discontinuous massless limit~\cite{vanDam:1970vg,Zakharov:1970cc} makes a qualitative difference from the vector QED case.\footnote{The authors would like to thank Zvi Bern for pointing this out.}
Another interesting future direction would be to elaborate on the Thomas-Wigner rotation factor associated with the definition of the polarisation tensor~\cite{Chung:2019duq,Chung:2020rrz}, which has an effect of switching from the covariant SSC to the canonical SSC, in a more general setup.

We conclude this paper with a curious observation regarding the gyromagnetic ratio and classical scattering amplitudes. The $(g-2)^2$ dependence of the difference \eqc{eq:CompClassDiff} implies that the overlap between physical external states and unphysical intermediate states is proportional to $(g-2)$, i.e. $ \la \psi_{\text{unphys}} | \psi_{\text{phys}} \ra \propto (g-2)$. Weinberg has commented that exponentiation of the eikonal amplitude for $g \neq 2$ holds when a sum rule constraining the anomalous magnetic moment $\D g = (g-2)$ to the amplitudes for radiative decay is satisfied~\cite{Weinberg:1971cdi}. The two statements are similar in that both are statements about $\D g$ of scattering amplitudes in the classical limit, and that both relate $\D g$ to transitions between observed states and unobserved states.

\vskip 1cm 

\acknowledgments
The authors appreciate Zvi Bern and Justin Vines for insightful discussions. JWK would like to thank Gustav Mogull and M.~V.~S. Saketh for helpful discussions, and Yilber Fabian Bautista for comments on the draft. JS is grateful for useful discussions with Maor Ben-Shahar.

\appendix
\section{Spin operator conversion for spinor and vector QED} \label{app:spinop}
For Dirac spinors, the following relations are used to reduce spinor products into spin tensor expressions.
\bl
\bld
\bar{u} (p_2+q) \slashed{A} u (p_2) &= \bar{u} (p_2+q) \frac{ \{ \slashed{A} , \slashed{p_2} \} + \slashed{q} \slashed{A}}{2m} u (p_2)
\\ &= \frac{(A \cdot p_2)}{m} \bar{u} (p_2+q) u (p_2) + \frac{1}{2m} \bar{u} (p_2+q) \slashed{q} \slashed{A} u (p_2)
\\ \bar{u} (p_2+q) \slashed{A} \slashed{B} u (p_2) &= \bar{u} (p_2+q) \frac{ \{ \slashed{A} , \slashed{B} \} + [ \slashed{A} , \slashed{B} ]}{2} u (p_2)
\\ &= (A \cdot B) \bar{u} (p_2+q) u (p_2) - 2 i A_\m B_\n \bar{u} (p_2+q) \mathbb{S}^{\m\n} u (p_2)
\eld
\el
Gamma matrix chains with more than three gamma matrices can be reduced to above cases using Clifford algebra $\{ \g^\m , \g^\n \} = 2 \eta^{\m\n}$.

For vector fields, the following decomposition of polarisation vector outer product~\cite{Chung:2019duq} is used to recast into spin vector expressions.
\bl
\ve^{\ast\m} (p) \ve^\n (p) &= - \left( \frac{\eta^{\m\n}}{2} - \frac{p^\m p^\n}{m^2} \right) \iden - \frac{i}{2m} \e^{\m\n\a\b} p_\a \mathbb{S}_\b - \left( \frac{\mathbb{S}^\m \mathbb{S}^\n + \mathbb{S}^\n \mathbb{S}^\m}{2} - \frac{\mathbb{S}_\a \mathbb{S}^\a \eta^{\m\n}}{4} \right) \,, \label{eq:Procapol2spin}
\el
where $\mathbb{S}_\a \mathbb{S}^\a = - \vec{\mathbb{S}} \cdot \vec{\mathbb{S}} = - 2 \hbar^2$ is the quadratic Casimir and $\iden$ is the identity operator in little group space. Unfortunately, this expression does not capture the quadratic Casimir dependence faithfully and following rules are adopted when interpreting the results.
\bn
\item If a term with $\hbar^{-2}$ scaling of the form $\hbar^{-2} A \mathbb{S}_\a \mathbb{S}^\a$ appears, it should be understood as $\hbar^{-2} A \mathbb{S}_\a \mathbb{S}^\a \to -2A \iden$ instead.
\item Since a classical term of dependence $q^2 \mathbb{S}^2$ cannot be distinguished from a quantum term of dependence $\hbar^2 q^2 \iden$, the results are analysed modulo $q^2 \mathbb{S}^2$ dependence.
\en
The polarisation vector for the outgoing particle needs to be rewritten in terms of the incoming particle momentum to use the relation \eqc{eq:Procapol2spin}. The following boost operation is used to relate the incoming polarisation vector $\ve_3^\m$ of the outgoing momentum $-p_3^\m$ to the outgoing polarisation vector $\ve_2^{\ast \m}$ of the incoming momentum $p_2^\m$.
\bl
\bld
\ve_3^\m &= G(-p_3;p_2)^\m_{~\n} \ve_2^{\ast\n} = \left[ \delta^\m_\n - \frac{(p_2 - p_3)^\m (p_2 - p_3)_\n}{m^2 + (-p_3 \cdot p_2)} + 2 \frac{(-p_3)^\m p_{2\n}}{m^2} \right] \ve_2^{\ast\n}
\\ &= \ve_2^{\ast\m} + 2 \frac{(p_2 - p_3)^\m (p_3 \cdot \ve_2^\ast)}{4m^2 - (p_2 + p_3)^2} \,.
\eld
\el
The Thomas-Wigner rotation factor~\cite{Chung:2019duq,Chung:2020rrz} is irrelevant for the Compton amplitude and has been neglected.

\section{Symplectic structure of Grassmann variables} \label{app:Grass_Quant}
Consider the following Lagrangian for the Grassmann variable $a$.
\bl
\CL = \frac{i}{2} a \dot{a} - V(a) \,.
\el
The canonical momentum is defined by a right derivative;
\bl
p_a := - \left. \frac{\delta \CL}{\delta \dot{a}} \right|_R = - \frac{i a}{2} \,.
\el
This means the momentum variable $p_a$ is \emph{not} independent from the position variable $a$, and this equation should be considered as a primary constraint $\phi_1$ which is weakly zero ($\phi_1 \simeq 0$).
\bl
\phi_1 := p_a + \frac{i a}{2} \simeq 0 \,.
\el
Due to the sign choice for $p_a$, the Hamiltonian is given as
\bl
\CH = - p_a \dot{a} - \CL = V(a) \,.
\el
The non-vanishing Poisson brackets $\{ \bullet , \bullet \}$ are given as
\bl
\{ a, p_a \} = + \{ p_a , a \} = 1 \,,
\el
where the sign factor is due to the graded nature of the algebra for Grassmann variables. Using Hamiltonian EOM $\{ f, \CH \} = \dot{f}$, the following secondary constraint is obtained.
\bl
\phi_2 = \dot{\phi_1} = \{ \phi_1 , \CH \} = \{ p_a , V(a) \} = \left. \frac{\partial V(a)}{\partial a} \right|_L \simeq 0 \,.
\el
The secondary constraint $\phi_2$ vanishes under the Poisson bracket since it is independent of $a$ and $p_a$,\footnote{A function $f(\eta)$ of a Grassmann variable $\eta$ can be at most linear in $\eta$.} which means that $\phi_2$ is a first-class constraint. On the other hand, the primary constraint $\phi_1$ has a nonvanishing Poisson bracket with itself, which means that $\phi_1$ is a second-class constraint;
\bl
\{ \phi_1 , \phi_1 \} = \{ p_a + \frac{ia}{2} , p_a + \frac{ia}{2} \} = i \,.
\el
This is troublesome since the points on the constraint surface $\phi_1 = 0$ get transported away from the constraint surface by the momentum map of the vector field $\{ \bullet , \phi_1 \}$, thereby making the constraint condition no longer preserved. Therefore the Poisson bracket is modified into the Dirac bracket to make $\phi_1$ vanish under the brackets. The Dirac bracket $\{ \bullet, \bullet \}^\ast$ is defined as;
\bl
\{ P, Q \}^\ast := \{ P, Q \} - \{ P, \phi_1 \} \left[ \{ \phi_1 , \phi_1 \} \right]^{-1} \{ \phi_1, Q \} = \{ P, Q \} + i \{ P, \phi_1 \} \{ \phi_1, Q \} \,.
\el
We find
\bl
\{ a, a \}^\ast = + i \,,\, \{ a, p_a \}^\ast = 1/2 \,,\, \{ p_a , p_a \}^\ast = - i / 4 \,,
\el
which is consistent with the constraint $\phi_1 = p_a + i a/2 \simeq 0$. The Poisson brackets in \eqc{eq:PB} for Grassmann variables should be understood as Dirac brackets.

\section{Fixing zeros of $h$ and $H$ parameters} \label{app:h0H0}
The $h=0$ case of the WQFT $\D g$ term \eqc{eq:delta_Ham} can be argued to be equivalent to the $H=0$ case of vector QED action \eqc{eq:spin1_action} by attempting to write it as a worldline action, following a procedure similar to that of ref.\cite{Bastianelli:2021nbs}. Up to total derivatives, the vector QED action for $H=0$ can be written as
\bl
\CL &= v^\dagger_\a \left[ (D_\m D^\m + m^2) \delta^\a_{\b} - \frac{g}{2} F_{\m\n} \left\{ i \eta^{\m\a} \delta^\n_\b - i \eta^{\n\a} \delta^\m_\b \right\} \right] v^\b + \cdots \,,
\el
where ellipsis denotes terms irrelevant for equations of motion of the field $v^\m$. Writing the covariant derivative as the mechanical momentum $D_\m = - i \pi_\m$ and the terms inside the curly brackets as the spin generator $\S^{\m\n}$, the action can be written as
\bl
\CL = - v^\dagger_\a \left[ (\pi_\m \pi^\m - m^2) \delta^\a_{\b} - \frac{ge}{2} F_{\m\n} \left(\S^{\m\n} \right)^\a_{~\b} \right] v^\b + \cdots \,, \label{eq:PQEDactForWQFT}
\el
where the charge of the particle $e = -1$ has been restored. The terms inside the square brackets of \eqc{eq:PQEDactForWQFT} matches the worldline Hamiltonian $\CH + \CH_{\D g,h}$ with $h=0$ up to an overall factor.
\bl
\bld
2 (\CH + \CH_{\D g,h=0}) &= \left( \pi^2 - m^2 - q F_{\m\n} S^{\m\n} \right) + \left( - \frac{\D g q}{2} F_{\m\n} S^{\m\n} \right)
\\ &= \pi^2 - m^2 - \frac{gq}{2} F_{\m\n} S^{\m\n} \,.
\eld
\el
It is unclear if this procedure of generating worldline actions from QFT actions can be generalised when projection conditions are imposed, e.g. for $H=1$ case of \eqc{eq:spin1_action}. In ref.\cite{Bastianelli:2021nbs} gauge-fixing terms were tuned to remove the projection terms of the Maxwell action.

\section{Classical equations of motion from the WQFT Hamiltonian}
The classical equations of motion can be obtained by substituting $S^{\m\n} = - i \psi^\m \psi^\n$ into the worldline Hamiltonian $\CH_{\text{tot}} = \CH + \CH_{\D g, h=1}$;
\bl
\CH_{\text{tot}} &= \CH + \CH_{\D g, h=1} = \half \left( \pi^2 - m^2 - \frac{gq}{2} F_{\m\n} S^{\m\n} - \D g q \frac{\pi^\a F_{\a\b} S^{\b\g} \pi_\g}{\pi^2} \right) \,,
\\ \dot{x}^\m &= \{ x^\m , \CH_{\text{tot}} \} = \pi^\m - \frac{\D g q S^{\m\a}F_{\a\b}\pi^\b}{2 \pi^2} + \CO(\pi_\a S^{\a\b}) \,,
\\ \dot{\pi}^\m &= \{ \pi^\m , \CH_{\text{tot}} \} = q F^{\m\a} \pi_\a + \frac{qg}{4} \partial^\m F_{\a\b} S^{\a\b} + \CO(F^2) + \CO(\pi_\a S^{\a\b}) \,,
\\ S^\m &= - \half \e^{\m\a\b\g} \pi_\a S_{\b\g} \,,\, S^{\m\n} = - \e^{\m\n\a\b} \pi_\a S_\b / \pi^2 \,,
\\ \dot{S}^\m &= \{ S^\m , \CH_{\text{tot}} \} = \frac{g q}{2} F^{\m\a} S_\a + \frac{\D g q}{2} \frac{\pi^\m S_\a F^{\a\b} \pi_\b}{\pi^2} + \CO(F^2) + \CO(S^2) + \CO(\pi_\a S^{\a\b}) \,,
\el
where $\CO(F^2)$ denotes nonlinear terms in the field strength, $\CO(\pi_\a S^{\a\b})$ denotes terms proportional to the covariant SSC, and $\CO(S^2)$ denotes terms quadratic in spin. Note that the last equation is the BMT equation~\cite{Bargmann:1959gz}. SSC is preserved due to $\{ \pi_\a S^{\a\b} , \CH_{\text{tot}} \} =  \CO(S^2) + \CO(\pi_\a S^{\a\b})$.

\bibliography{SSCinQFT,mybib}
\bibliographystyle{JHEP}
\end{document}